\documentstyle[12pt,aasms4]{article}
\def\lapprox{\hbox{\lower .8ex\hbox{$\,\buildrel < \over\sim\,$}}}
\def\gapprox{\hbox{\lower .8ex\hbox{$\,\buildrel > \over\sim\,$}}}

\begin{document}

\bigskip
\bigskip
\bigskip
\bigskip

\title
{The cosmic rate of supernovae and the range of stars ending as Type Ia SNe}

\author
{P. Ruiz--Lapuente \altaffilmark{1,2} \& R. Canal \altaffilmark{1}}

\altaffiltext
{1}{ Department of Astronomy, University of Barcelona, Mart\'{\i}\ i 
Franqu\'es 1, E--08028 Barcelona, Spain. E--mail: pilar@mizar.am.ub.es,
ramon@mizar.am.ub.es}

\altaffiltext
{2}{Max--Planck--Institut f\"ur Astrophysik, Karl--Schwarzschild--Strasse 1,
D--85740 Garching, Federal Republic of Germany. E--mail:
pilar@MPA--Garching.MPG.DE}

\slugcomment{{\it Running title:} The cosmic rate of supernovae}

\begin{abstract}
The present cosmic rate of Type Ia supernovae (SNeIa) suggests that about 
6\% of all stars in binary systems with primaries in the initial mass range 
$3-9\ M\sun$ end up as SNeIa. If that is confirmed, the unavoidable 
conclusion is that SNeIa can only be explained by the single degenerate 
scenario. At most 1\% of stars in binary systems in the above range end up as 
CO + CO WD pairs, with total mass equal to or larger than the Chandrasekhar 
mass. Given that the number of mergers from pairs of CO + He WDs that reach 
the Chandrasekhar mass is even lower, the conclusion strongly favors binaries 
containing just one CO WD as the progenitors of SNeIa, since the SNeIa 
production efficiency (relative to the instantaneous star formation rate)
predicted for double degenerate (DD) pairs lies more than $3\sigma$ below 
the observational data, and the DD scenario can be rejected at more than 99\% 
confidence level. Only if the SFR measurements from $z\sim 0.1$ to 
$z\sim 0.5$ are being underestimated by a factor of 6 while SNeIa rates are 
not, can we escape the above conclusion. We evaluate the numbers and 
characteristics of double WD systems with different chemical compositions 
(CO and He WDs) that should form and compare them with the observations, in 
order to check our predictions. Our conclusions appear robust after that 
test. 
\end{abstract}


\section{Introduction}

The evolution of intermediate and low--mass stars in binary systems is a 
complex subject that has, however, enormous importance in astrophysics. Of 
their way to the final end, a bulk of information is available from 
observations on the processes undergone by binaries as they evolve, i.e. 
common--envelope phases, accretion, outburst.

A relevant new piece of information comes from the consideration of the 
number of Type Ia supernova explosions (SNeIa) that occur per unit mass spent 
in forming stars in different environments. So far, the exploration of the 
issue was restricted to nearby galaxies, where the combination of SNeIa 
statistics with star formation rate (SFR) history is poorly known.

The approach explored here is to take the average of star formation history 
in redshift space and use the information on cosmic SNeIa rate to infer how 
many stars in binaries from low to intermediate mass end up as SNeIa. This 
indirect approach started a few years ago (Ruiz--Lapuente, Burkert, \& Canal 
1995; Ruiz--Lapuente, Canal, \& Burkert 1997; see also Madau, Della Valle, 
\& Panagia 1998; Sadat et al. 1998; Ruiz--Lapuente \& Canal 1998; Yungelson 
\& Livio 1998, 2000; Kobayashi et al. 1998; Dahlen \& Fransson 1999; 
Ruiz--Lapuente, Cass\'e, \& Vangioni--Flam 2000) and it can help us to 
determine the nature of the progenitors of SNeIa.

We evaluate the number of SNeIa exploding per unit comoving volume, in 
redshift space, and relative to the mass going into forming stars in the 
Universe. Despite the uncertainties still involved, the last years have 
nevertheless brought some crucial information.

As we will see, if the efficiency of binaries in ending as SNeIa is 6\% of 
the stars between $3-9\ M\sun$, no other alternative than the single 
degenerate scenario appears as a reasonable candidate to explain those
explosions. The modeling on which that conclusion is based stands comparison 
with the statistics of the white dwarf (WD) pairs actually observed.

\section{The star formation rate and the SNeIa production efficiency}

In the last few years we have started to learn about the cosmic history of
the star formation process, and several groups (Madau et al. 1996; Steidel
et al. 1998; Hughes et al. 1998; Madau, Pozzetti, \& Dickinson 1998; Blain 
et al. 1999) have begun to derive the global star formation rate 
$\dot \rho_{*}(z)$. At the same time, the first measurements of the global 
SNeIa rates, $\Re_{Ia}(z)$, are being performed (Pain et al. 1996; Hardin et 
al. 2000; Hamilton 1999; Hamuy \& Pinto 1999; Pain et al. 2000). The SNeIa 
rate is doubly related to the SFR: through the time delay between formation 
of the SNeIa progenitor systems and explosion, and through the fraction of 
stars (binaries in this case) that give rise to SNeIa. The comparison between 
$\dot \rho_{*}(z)$ and $\Re_{Ia}(z)$ thus contains key information on the 
nature of the so far elusive SNeIa progenitors. 

The work of Madau et al. (1996), based on the UV luminosities of galaxies in 
the Hubble Deep Field, complemented that of Lilly et al. (1995, 1996) at 
optical wavelengths in showing that the star formation activity steadily 
increases with $z$ from the local, present--day Universe, up to $z\sim 1.5$. 
The original claim that $\dot \rho_{*}(z)$ peaks there to fall again at 
higher $z$ has later been revised, in particular on basis to data at long 
wavelengths that would indicate that the star formation regions at high $z$ 
are enshrouded in dust (see Hughes et al. 1998, for instance). 
$\dot \rho_{*}(z)$ would level--off beyond $z\simeq 1.5-2$, up to much higher 
redshifts. Observations in the submillimeter range (Blain et al. 1999), ISO 
measurements of the extragalactic background light (Rowan--Robinson et al. 
1997; Flores et al. 1998; Elbaz et al. 1999), and the FIRAS and DIRBE 
experiments on board of COBE (Dwek et al. 1998; Fixsen et al. 1998), do in 
principle lift the veil on the star formation activity in dust--obscured 
regions, but there is no redshift identification of the emission detected. 
Measurement of the SN rates at high $z$ then would help to trace the star 
formation history. SNeIa are the most luminous ones. Thus, knowledge of the 
characteristic time delays between formation of their progenitor systems and 
explosion is most relevant.

The measurements of the cosmic evolution of the SNeIa rate now extend up to 
$z\sim 0.55$, and there are already preliminary results for $z\simeq 1.1$.  
Current results are summarized in Table 1. By combining our knowledge of 
$\Re_{Ia}(z)$ with that of $\dot \rho_{*}(z)$, we can now gain new insight on 
the nature of the SNeIa progenitor systems. We introduce for that the 
``efficiency'' ${\cal E}_{SNeIa}(z)$ of SNeIa production referred to the SFR, 
both per unit of comoving volume:

$${\cal E}_{SNeIa}(z) = \Re_{Ia}(z)\ yr^{-1}\ M\!pc^{-3}\ /\ 
\dot \rho_{*}(z)\ M_{\odot}\ yr^{-1}\ M\!pc^{-3}\eqno(1)$$

The quantity ${\cal E}_{SNeIa}(z)$ (given in the fourth column of 
Table 1) is thus just the number of SNeIa per unit mass spent in forming 
stars at a given $z$, and it is independent from the cosmological model 
assumed. This efficiency reflects the evolutionary time scale (from birth to 
explosion) of the progenitor systems and the range of initial conditions 
leading to SNeIa explosions, together with other possible evolutionary 
effects (such as dependence on initial metallicity). The dependence on the 
time scale is illustrated in Figure 1 for the two main types of progenitor 
systems so far proposed, and for three different $\dot \rho_{*}(z)$. The 
systems labelled CLS(W) have evolutionary time scales of the order of a few 
Gyr whereas those labelled DD have time scales of the order of a few hundred 
million years only. The ``flat'' evolution of the efficiency for the DD 
systems for decreasing $z$ is due to the fact that $\Re_{Ia}(z)$ closely 
follows $\dot \rho_{*}(z)$. Instead, ${\cal E}_{SNeIa}(z)$ significantly 
increases towards lower $z$ for the CLS(W) systems because the systems now 
exploding are a fixed fraction of the stars formed a few Gyr ago, when 
$\dot \rho_{*}(z)$ was higher. The flattening of  ${\cal E}_{SNeIa}(z)$ at 
larger $z$ corresponds to a similar flattening of the $\dot \rho_{*}(z)$ 
considered. The absolute values of ${\cal E}_{SNeIa}(z)$ reflect the 
abundance of progenitor systems (and thus the range of initial conditions 
leading to SNeIa), and they will be discussed below.

The two types of systems whose ${\cal E}_{SNeIa}(z)$ is shown in Figure 1 have 
emerged in the last years as the main candidates to SNeIa progenitors.
The DD systems are binaries made of a couple of CO WDs with a total mass
exceeding the Chandrasekhar mass, and close enough to merge due to emission 
of gravitational waves in less than a Hubble time (Webbink 1984; Iben and 
Tutukov 1984). The CLS(W) systems consist of a CO WD plus a subgiant or
red--giant companion that is overfilling its Roche lobe and transferring
matter from its envelope to the WD ({\it cataclysmic--like system} or 
Algol--type binary that might be observed as a {\it supersoft X--ray source}). 
This scenario was originally proposed by Whelan \& Iben (1973) and recently 
refined by Hachisu, Kato, \& Nomoto (1996), who introduce the effects of 
a strong wind emitted by the accreting WD to stabilize mass transfer. That
would allow the WD to reach the Chandrasekhar mass by steadily
burning H into He and then He into CO (see also Nomoto, Iwamoto, \& Kishimoto
1997). General discussions of the possible SNeIa progenitor systems can be
found in Ruiz--Lapuente, Canal, \& Burkert (1997).

The ${\cal E}_{SNeIa}(z)$ values in Table 1 are based on $\dot \rho_{*} = 
2.01^{+0.18}_{-0.18}\times 10^{-2}\ h_{65}\ M_{\odot}\ M\!pc^{-3}\ yr^{-1}$ 
(Gronwall 1999) for the local SFR. For higher $z$, we give the average of the 
values calculated  for each of the three different $\dot \rho_{*}(z)$ 
referred to in Figure 1. The low--redshift efficiencies correspond to 1 SNeIa 
per $\sim 900 M_{\odot}$ going into star formation. As we see, 
${\cal E}_{SNeIa}(z)$ is of the same order up to $z\sim 0.55$. Ongoing 
searches will soon yield the efficiency at $z\simeq 1$. If we assume that all 
stars with initial masses $M\gapprox 10\ M_{\odot}$ produce SNeII + SNeIb/c 
(gravitational--collapse SN), then $\Re_{II+Ib/c}(z)\simeq 0.0054\times \dot 
\rho_{*}(z)$, for a Salpeter initial mass function (IMF) with $x = 1.35$, 
extending from 0.1 M$_{\odot}$ up to 100 M$_{\odot}$. With the above values, 
that gives about 5 gravitational--collapse SN for every SNeIa in the local 
Universe (see Table1), in good agreement with the observations.

We can relate ${\cal E}_{SNeIa}(z)$ to the fraction $\eta$ of stars in the 
$3\ M_{\odot}\lapprox M\lapprox 9\ M_{\odot}$ initial mass range that should
produce SNeIa if $\Re_{Ia}(z)$ closely followed $\dot \rho_{*}(z)$, since we 
would then have:

$${\cal E}_{SNeIa}(z) = \eta \times {{\int_{3}^{9} \Phi (M)\ dM}\over
{\int_{0.1}^{100} M\ \Phi(M)\ dM}}\eqno(2)$$

\noindent
where $\Phi (M)$ is the IMF. For the same IMF as above, we would have 
${\cal E}_{SNeIa}(z) =  0.02304\times \eta$, which would mean 
$\eta\simeq 0.06$ for the $\langle{\cal E}_{SNeIa}\rangle$ given below. 
Some $6\%$ of the stars in the above mass range (making the approximation 
that all stars were born in binaries) should thus give rise to SNeIa. Given 
the strong selection that any initial population of binaries with primaries 
in that mass range undergoes in any scenario, at different evolutionary 
steps, until becoming a SNeIa candidate, the figure looks high if there is no 
help from the fact that most systems now exploding were formed at an epoch 
when $\dot \rho_{*}(z)$ was considerably higher than at the redshifts 
considered here, and it confirms the impression carried by Figure 1 that the 
DD systems, with their short evolutionary timescales adding to the strong 
evolutionary constraints, do fail to account for the bulk of observed SNeIa. 
We will see in the next Section that modeling of the Galactic DD population 
reinforces this conclusion. Comparing the measured ${\cal E}_{SNeIa}$ with 
that predicted from the DD scenario, at $z\simeq 0.55$ for instance, one sees 
that the prediction (${\cal E}_{SNeIa} = 1.80\times 10^{-4}\ M_{\odot}^{-1}$) 
is more than $2.95\sigma$ below the measurement (${\cal E}_{SNeIa} = 
1.42 ^{+0.45}_{-0.42}\times 10^{-3}\ M_{\odot}^{-1}$). The $\chi^{2}$ test 
gives a probability $P < 0.006$ that the two efficiencies were the same. Even 
allowing for a sytematic error by a factor of 2 in the theoretical prediction 
of the rates and taking then the upper limit, the prediction would still be 
more than $2.5\sigma$ below the observational value ($P < 0.012$). If we refer 
to average values over the interval $0\leq z\leq 0.55$, then we have, from 
observations, $\langle{\cal E}_{SNeIa}\rangle = 1.41 
^{+0.40}_{-0.31}\times 10^{-3}\ M_{\odot}^{-1}$ whereas the DD prediction is
$2.17\times 10^{-4}\ M_{\odot}^{-1}$, more than $3.8\sigma$ below 
($P < 0.004$). 

Precise enough measurements of $\Re_{Ia}(z)$ at $z\gapprox 1$ would in turn
test the prediction that at those redshifts the decrease in metallicity
should inhibit the strong wind mechanism in the CLS(W) systems (Kobayashi et
al. 1998). A decrease in ${\cal E}_{SNeIa}(z)$ should be observed, even if
SNeIa do not completely disappear thanks to the early metal enrichment of
elliptical galaxies.

\subsection{The local density of double degenerate binaries}

The accuracy of model predictions of the SNeIa rates expected from the merging 
of two CO WDs (the DD scenario) cannot just be tested against the actually 
measured rates, since the observed SNeIa can arise from a different 
evolutionary pathway. Only predicted rates much in excess of the measured ones 
would prove the modeling wrong. Since we find the efficency of SNeIa 
production in the DD scenario to be too low as compared with that observed, 
some extra test of the general binary evolution model is required. Such a 
test is provided by the measured space density of binary systems consisting 
of a pair of WDs (DD systems). In the DD scenario, a fraction of those 
systems (close CO WD pairs with a total mass exceeding the Chandrasekhar mass 
and close enough to merge in less than a Hubble time) would be the SNeIa 
progenitors. 

From existing surveys, no clear SNeIa progenitor candidate has been found 
among the detected DD systems, but aside from the role played by not too well 
controlled selection effects, the fact does not automatically validate model 
predictions that the space density of possible progenitors should be low 
enough for them to have escaped detection up to now. A better test is 
provided by the comparison of model predictions with the sample of DD systems 
already detected, even if no SNeIa candidate has yet been found. We have thus 
calculated, with the same Monte Carlo scenario code used to predict the SNeIa 
rates for different evolutionary scenarios (see Ruiz--Lapuente, Canal, \& 
Burkert 1997, for instance), the numbers of different types of DD binaries 
(He WD + He WD, CO WD + He WD, and CO WD + CO WD) that would form at times 
following an instantaneous burst of star formation. The formation of WD + MS 
pairs is also followed. That is later convolved with a SFR appropriate for 
the Solar neighbourhood (constant rate, $\sim 1\ M_{\odot}\ yr^{-1}$ for the 
whole Galactic disk), which gives us the number density of DD systems 
existing at the present time. Since WD pairs older than $\sim 10^{8}\ yr$ 
should haved cooled below detection limits, the local SFR history is not 
really important when comparing the model with the observations. We find that 
the space density of WDs with ages below $10^{9}\ yr$ (wich roughly 
corresponds to $M_{V}\leq 12.75$) should be $\sim 2.5\times 10^{-3}\ 
pc^{-3}$, which agrees well with the observational estimates of Liebert, 
Dahn, \& Monet (1988) ($\sim 3\times 10^{-3}\ pc^{-3}$).

In Table 2 (first line) we give the predicted fractions of close DD and close 
WD + MS pairs versus all WDs (single plus both wide and close pairs). We also 
give the fraction corresponding to the predicted SNeIa candidates. In the 
second line is shown the current birth rate of the three types of systems, 
and in the third one the numbers of systems with ages below $10^{8}\ yr$. In 
the calculations we have adopted a value of the common envelope parameter 
$\alpha_{CE} = 1$. Our results compare well with the observed samples (Saffer,
Livio, \& Yungelson 1998; Maxted \& Marsh 1999). The predicted DD period and 
mass distributions do also agree (see Figure 2), especially if we take into 
account the observational biases against detection of systems with very low 
mass primaries (taken here to be the  brightest, and thus the more recently 
formed members of the DD systems), and of binaries with either the longest or 
the shortest periods. Therefore, our modeling of the possible DD progenitors 
of SNeIa can be considered as being tested against all the currently 
available observational data, and our conclusion as to the low efficiency of 
the DD scenario in producing possible SNeIa progenitors, as compared with the 
measurements of the cosmic evolution of the SNeIa rates, appears robust.

\section{Discussion and conclusions}

We have shown how, by using a cosmic approach, the topic of which stars 
end their evolution as SNeIa can be enlightened. About 6\% of the 
binary stars in the range $3-9\ M\sun$ end their lives as SNeIa. This 
fraction is unattainable from the merging of C+O WD pairs, since only about 
1\% of binaries with masses in that range have a final total mass which is 
above the Chandrasekhar mass. The high production efficiency of SNeIa out of 
the star formation process found from $z\sim 0$ to $z\sim 0.5$ and beyond 
suggests that the single degenerate scenario best explains Type Ia explosions. 

\noindent 
The above conclusion can be avoided if

\noindent
1. The star formation rate is understestimated by a factor of 6 even in the 
reconstructions from $z\sim 0$ till $z\sim 0.5$  which give the highest 
estimates, while the rate of SNeIa is not underestimated. This situation 
seems unlikely as it would imply that the real SFR history is inconsistent 
with a dozen of empirical determination using very different methods. In 
addition, such extremely high SFRs would be in conflict with all we
know about extragalactic background light. More especifically, at $z = 0.55$ 
there is just a discrepancy by a factor $\lapprox 1.5$ among the SFRs in 
Figure 1. The bulk of DD systems merging at that $z$ were formed at 
$z\simeq 0.62$, where the range of possible SFRs is not any broader.

\noindent
2. We have overestimated by a factor of 6 the empirical rates of SNeIa  from 
$z\sim 0$ till  $z\sim 0.5$. The possibility of having 
overestimated the SNeIa rate is unlikely since critical aspects 
such as obscuration by dust, or undetectability of SNeIa near galactic 
nuclei, should produce the opposite effect.

\clearpage

\centerline{\bf Tables}

\bigskip
\bigskip
\bigskip
\bigskip
\bigskip
\bigskip

\begin{table*}[htb]
\caption{Type Ia supernova rates and production efficiencies along z}
\label{table:1}
\newcommand{\m}{\hphantom{$-$}}
\newcommand{\cc}[1]{\multicolumn{1}{c}{#1}}
\renewcommand{\arraystretch}{1.2} % enlarge line spacing
\begin{tabular}{@{}lllll}
\hline
Redshift  & $\tau_{SNu}$ & $\Re_{Ia}$  & ${\cal E}_{SNeIa}$ & Probability \\

$\langle z\rangle$  & SNu  h$_{65}^{2}$ & Mpc$^{-3}$yr$^{-1}$h$_{65}^{3}$ &
 M$_{\odot}^{-1}$h$_{65}^{2}$ &  ($\chi^{2}$ test) \\
\hline
 0.$^{1}$ & 0.21$^{+0.30}_{-0.13}$ & 2.2$^{+3.4}_{-1.4}$ 10$^{-5}$ &  
 1.09$^{+1.55}_{-0.64}$ 10$^{-3}$ $^{**}$ &   \\
 0.$^{2}$ & 0.16$\pm$0.05 &  ---    &  ---  &  \\ 
 0.1$^{3}$ & 0.12$^{+0.13}_{-0.08}$ &  1.7$^{+1.9}_{-1.1}$ 10$^{-5}$  &
 1.24$^{+1.39}_{-0.81}$ 10$^{-3}$ $^{***}$ &   \\
 0.32$^{4}$ &  $<$ 0.32 (1$\sigma$) & $<$ 4.52 (1$\sigma$) 11.02
 (2$\sigma$) 10$^{-5}$  
 & $<$ 2.27 -- 5.54  10$^{-3}$ $^{***}$ &     \\
     &                  & $<$ 6.2\ \ \ (1$\sigma$) 15.0\ \ (2$\sigma$)
  10$^{-5}$ $^{*}$  & 
                      &    \\
 0.38$^{5}$ & 0.35$^{+0.38}_{-0.26}$ & 4.8$^{+3.3}_{-2.2}$ 10$^{-5}$  & 
 2.20$^{+1.51}_{-1.01}$ 10$^{-3}$ $^{***}$      &   \\
     &                & 6.9$^{+4.8}_{-3.2}$ 10$^{-5}$ $^{*}$  & 
                      &    \\
 0.55$^{6}$ & 0.26$\pm$ 0.08 & 4.53$^{+ 1.43}_{-1.35}$ 10$^{-5}$  &
 1.42$^{+0.45}_{-0.42}$ 10$^{-3}$ $^{***}$      &    \\
        &    &6.74$^{+2.13}_{-2.00}$  10$^{-5}$ $^{*} $ &
      &        \\
        &    &                                          &
      & P $<$ 0.004 \\
\hline
\end{tabular}\\[2pt]
$^{1}$Hamuy \& Pinto (1999); $^{2}$Cappellaro et al. (1997); 
$^{3}$Hardin et al. (1999); $^{4}$Hamilton (1999); $^{5}$Pain et al.(1996);
$^{6}$Pain et al.(2000) 
$^{*}$For the cosmology $\Omega_{M}$ =0.3 $\Omega_{\Lambda}$=0.0;
$^{**}$SFR from Gronwall (1999);
$^{***}$Averaging over the SFRs used in Figure 1;   
\end{table*}

\begin{table*}[htb]
\caption{Predicted fractions of DDs and of WD + MS pairs}
\label{table:2}
\newcommand{\m}{\hphantom{$-$}}
\newcommand{\cc}[1]{\multicolumn{1}{c}{#1}}
\renewcommand{\arraystretch}{1.2} % enlarge line spacing
\begin{tabular}{@{}llllll}
\hline

Close DD & CO + CO & Pre--SNeIa & CO + He & He + He & Close WD + MS \\

\hline

1/15  & 1/114 & 1/419 & 1/24  & 1/58  & 1/6   \\

 0.056 & 0.008 & 0.002 & 0.037 & 0.011 & 0.146 \\

5.6$\times 10^{6}$ & 8.5$\times 10^{5}$ & 2.2$\times 10^{5}$ & 
3.7$\times 10^{6}$ & 1.1$\times 10^{6}$ & 1.5$\times 10^{7}$  \\  

\hline

\end{tabular}
\end{table*}

\clearpage

\begin{figure}[hbtp]
\centerline{\epsfysize16cm\epsfbox{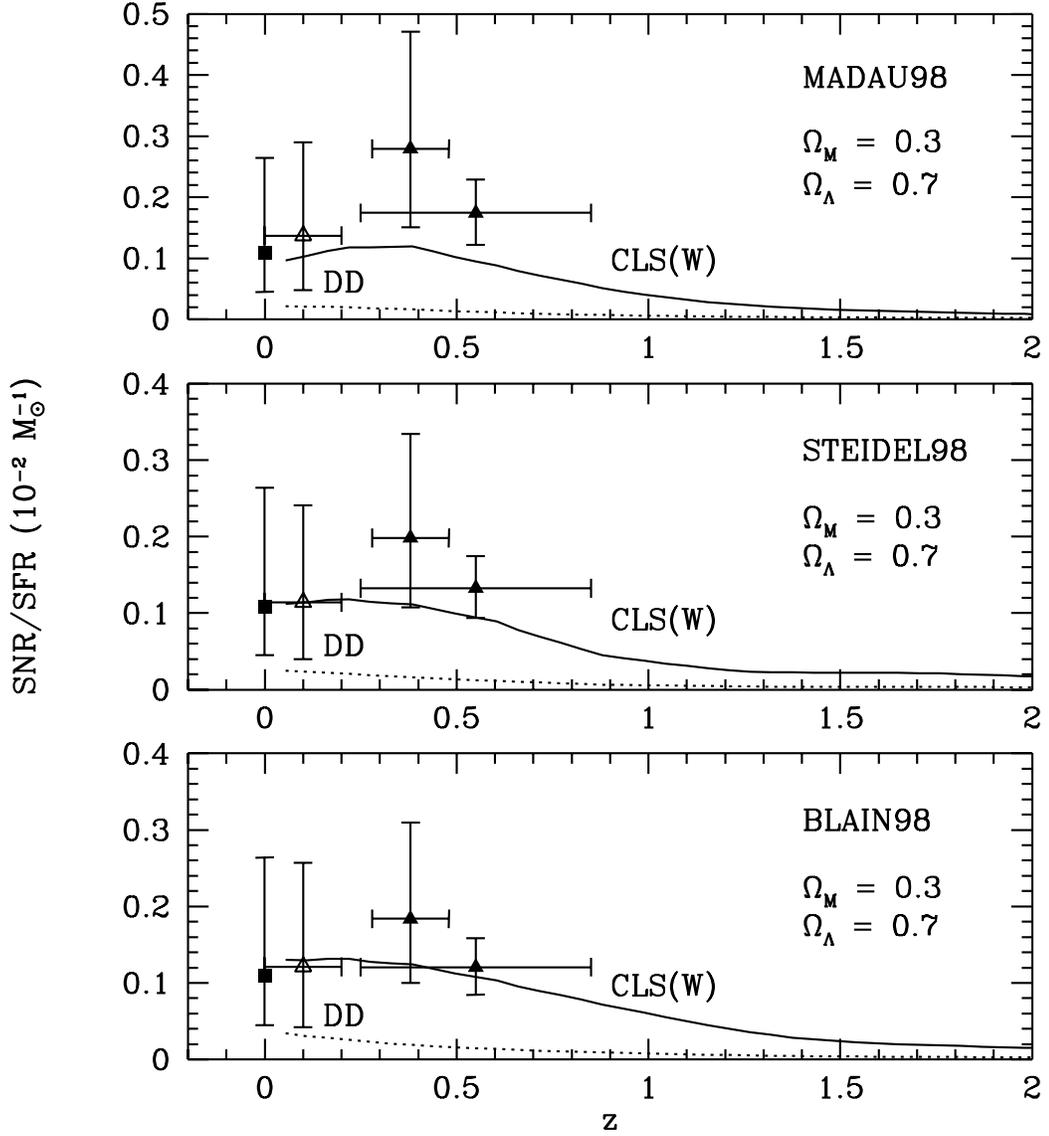}}

\nopagebreak[4]

\figcaption{Model predictions of ${\cal E}_{SNeIa}$, the ``efficiency'' in 
producing SNeIa, that is the number of SN per unit mass of stars being 
formed, at a given $z$. The curves show its expected evolution for two SNeIa 
candidates systems with different timescales to explosion: merging of two CO 
WDs (dashed line, labelled DD), and cataclysmic--like binaries (or 
Algol--type systems), with the stabilizing effects of the accretion--induced 
wind included (solid line, labelled CLS(W), see text for further 
explanation). The results for three different star--formation histories are 
shown: Madau et al. (1998), Steidel et al. (1998), and Blain et al. (1998). 
The data points come from the SNeIa rate measurements of Pain et al. (1996, 
2000), Hardin et al. (1999), and Hamuy \& Pinto (1999). The results are 
independent from the cosmological model adopted. Note that the vertical scale
of the first panel is different from the other two.}
\label{fig1}
\end{figure}

\clearpage

\begin{figure}[hbtp]
\centerline{\epsfysize16cm\epsfbox{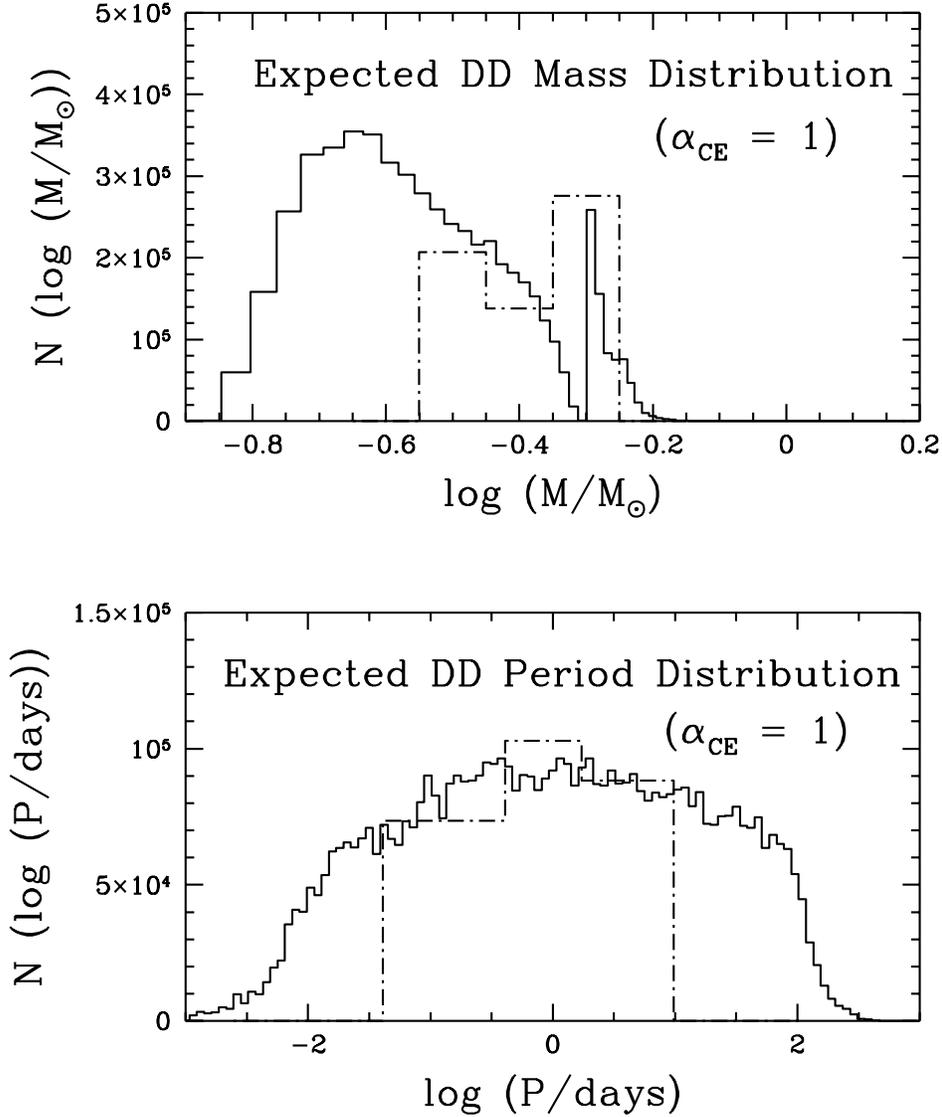}}

\nopagebreak[4]

\figcaption{{\it Top panel}: Expected mass distribution of the brightest 
components of DD systems (solid histogram). The dot--dashed histogram shows 
the distribution in the observed sample. {\it Bottom panel}: Expected period 
distribution of DD systems (solid histogram). The dot--dashed histogram shows 
the distribution in the observed sample. The scarcity of low--mass WDs in the 
observed sample, as compared with the model prediction, can be explained by
the increasing difficulty in detecting lower mass WDs, and the same applies
to either very short or very long periods.
 } 
\label{fig2}
\end{figure}

\end{document}